
\magnification=1200
\tolerance=10000
\font\titlea=cmb10 scaled\magstep1
\font\titleb=cmb10 scaled\magstep2
\rightline{UG-5/94}
\rightline{hep-th/9406131}
\baselineskip=18pt
\vskip 1cm
\centerline{\titleb Observations on the Topological Structure}
\centerline{\titleb in 2d Gravity Coupled to Minimal Matter}
\vskip 2cm
\baselineskip=14pt
\centerline{Pablo M. Llatas\footnote{$^1$}{e-mail address:
Llatas@th.rug.nl}}
\medskip
\centerline{and}
\medskip
\centerline{Shibaji Roy\footnote{$^2$}{e-mail address: Roy@th.rug.nl}}
\bigskip
\centerline{\it Institute for Theoretical Physics}
\centerline{\it Nijenborgh 4, 9747 AG Groningen}
\centerline{\it The Netherlands}
\vskip 1cm
\baselineskip=18pt
\centerline{\titlea ABSTRACT}
\bigskip
By using a bosonization we uncover the topological gravity structure of
Labastida, Pernici and Witten in ordinary $2d$ gravity coupled to
$(p,q)$ minimal models. We study the cohomology class associated
with the fermionic charge of the topological gravity which is shown
to be isomorphic to that of the total $BRST$ charge. One
of the ground ring
generators of $c_M <1$ string theory is found to be in the equivariant
cohomology of this fermionic charge.
\vfill
\eject

It is by now fairly well understood [1,2] that $(q-1)$
(where $q\geq 2$) matrix
model at various multicritical points is equivalent to $2d$ topological
gravity coupled to twisted $N=2$ superconformal matter field theory [3,4] with
central charge $c^{N=2} ={3(q-2)\over q}$. In order to understand the
relations between these theories and ordinary $2d$ gravity coupled to
$(p,q)$ minimal models, it is desirable to understand the topological
structure of the latter. In the context of one-matrix model, Distler [5] has
shown that the multicritical points can be reached from the base topological
field theory by using marginal perturbation. In this case the basic
topological theory is the $2d$ topological gravity of Labastida, Pernici
and Witten [6]. $2d$ topological
gravity in the formulation of Distler has been
understood as $c_M =-2$ minimal matter or $(1,2)$ minimal model coupled to
Liouville gravity. Another formulation of topological gravity has been
described in ref. [7,8]. More recently, a series of topological structures
with varying central charges have been revealed [9-11] in $2d$ gravity coupled
to $(p,q)$ minimal matter.

In this paper, we describe another topological structure namely, the
topological gravity structure of Labastida, Pernici and Witten in ordinary
$2d$ gravity coupled to all $(p,q)$ minimal models. We use a specific
bosonization (depending on $(p,q)$) of the Liouville and matter fields
of the $c_M <1$ string theory in terms of the bosonic fields $\beta (z)$
and $\gamma (z)$ of the topological gravity. Under this bosonization
the energy-momentum tensor of $c_M <1$ string theory reduces to that
of the topological gravity. Also, for $p=1$ and $q=2$, our bosonization
reduces to that found by Distler [5]. In analogy with topological gravity,
we show that there is a twisted $N=2$ superconformal algebra with
central charge $c^{N=2} =-9$ in $2d$ gravity coupled to $(p,q)$
minimal matter. By using a similarity transformation we show that the
cohomology class associated with the total $BRST$ charge is isomorphic
to that of the fermionic charge $Q_s$ of the topological gravity. We
study the cohomology class associated with this fermionic charge. In
particular, we find that one of the ground ring generators [12,13] can be
expressed in a local way in terms of the fields of the topological
gravity, while the other can not. We find them both in the equivariant
cohomology of $Q_s$.

The total energy-momentum tensor of $(p,q)$ minimal model
coupled to $2d$ gravity (where $gcd~(p,q)=1$) is

$$
T(z) = T_M(z) + T_L(z) + T^{gh}(z)\eqno(1)
$$
where
$$
\eqalign{T_M(z) &= -{1\over 2} :\partial\phi_M(z)\partial\phi_M(z):
+ i Q_M \partial^2\phi_M(z)\cr
T_L(z) &= -{1\over 2} :\partial\phi_L(z)\partial\phi_L(z): + i Q_L
\partial^2\phi_L(z)\cr
T^{gh}(z) &= -2 :b(z)\partial c(z): - :\partial b(z) c(z):\cr}\eqno(2)
$$
are the energy-momentum tensors of the matter, Liouville and the ghost
sectors with $\phi_M(z)$, $\phi_L(z)$, the matter and Liouville fields and
$(b(z),c(z))$,
the reparametrization ghost system having conformal weights 2 and $-1$
respectively. $2 Q_M$ and $2 Q_L$ are the background charges of the
matter and Liouville sector. Since the total central charge of the
combined matter-Liouville system is 26, they satisfy
$Q_M^2 + Q_L^2 = -2$, where the basic OPE's are given as
$<\phi_M(z)\phi_M(w)>=<\phi_L(z)\phi_L(w)>=-\ln (z-w)$ and
$<b(z)c(w)>={1\over (z-w)}$. For any $\lambda >0$, the background charges
can be parametrized as,

$$
\eqalign{Q_L &= i (\lambda + {1\over {2\lambda}})\cr
Q_M &=({1\over {2\lambda}} - \lambda)}\eqno(3)
$$

We now introduce two more conformal fields, $:e^{\pm i\lambda (\phi_M(z) -
i\phi_L(z))}:$ with conformal weight $\mp 1$ with respect to (1). Note that
for $(p,q)$ minimal model $\lambda =\sqrt{q\over{2p}}$ and also, that the
energy-momentum tensor (1) has the symmetries $\lambda\leftrightarrow
{1\over{2\lambda}}$, $\phi_M(z)\leftrightarrow -\phi_M(z)$ and
$\lambda\leftrightarrow {1\over{2i\lambda}}$, $\phi_M(z)\leftrightarrow
\phi_L(z)$.
Let us introduce two bosonic fields with conformal weight $-1$ and 2, defined
as:

$$
\eqalign{\gamma (z) &=:e^{i\lambda (\phi_M(z)-i\phi_L(z))}:\cr
\beta (z) &={1\over 2}:\biggl[(\lambda -{1\over{\lambda}})\partial\phi_L(z)
+i(\lambda +{1\over{\lambda}})\partial\phi_M(z)\biggr]
\ e^{-i\lambda (\phi_M(z)-i\phi_L(z))}:}\eqno(4)
$$
They satisfy the following non-vanishing basic OPE's:
$$
\beta (z)\gamma (w)= -\gamma (z)\beta (w)\sim  {1\over (z-w)}\eqno(5)
$$
In terms of these fields, the energy-momentum tensor (1) can be written
as:
$$
T(z)=2:\beta (z)\partial\gamma (z):+:\gamma (z)\partial\beta (z):-2:b(z)
\partial c(z):-:\partial b(z) c(z):\eqno(6)
$$

The total central charge associated with (6) is zero, since the bosonic
$(\beta (z) ,\gamma (z) )$
part contributes +26 and the fermionic $(b(z),c(z))$ part
contributes $-$26 to the central charge. The energy-momentum tensor (6)
can be derived from the topological gravity action considered by Labastida,
Pernici and Witten [6]. We also note that $\gamma (z)$ and $\beta (z)$ reduce
precisely to the form of bosonization considered by Distler [5] for
$\lambda =1$, $i.e.$, for $(1,2)$ model coupled to $2d$ gravity.

In another formulation of topological gravity [7], an auxiliary Liouville
sector was considered apart from the $(b(z),c(z),
\beta (z) ,\gamma (z) )$ sector.
But since the Liouville sector, as it is mentioned there, does not arise
from any conformal anomaly, it is not necessary to include them. In that
context, it was found that the topological gravity contains a twisted
$N=2$ superconformal algebra. In analogy, we also find a topological
conformal algebra with $c^{N=2}=-9$ in $2d$ gravity coupled to $(p,q)$
minimal model. In fact, the generator (6) along with
$$
\eqalignno{Q_s (z)&=b(z)\gamma (z) &(7a)\cr
J(z) &=:c(z)b(z):+2:\beta (z)\gamma (z): &(7b)\cr
G(z) &=c(z)\partial\beta (z)+2\partial c(z)\beta (z) &(7c)}
$$
satisfy the following twisted $N=2$ superconformal algebra:
$$
\eqalign{T(z)T(w) &\sim {2T(w)\over (z-w)^2}+{\partial T(w)\over (z-w)}\cr
T(z)Q_s(w) &\sim {Q_s(w)\over (z-w)^2}+{\partial Q_s(w)\over (z-w)}\cr
T(z)G(w) &\sim {2G(w)\over (z-w)^2}+{\partial G(w)\over (z-w)}\cr
T(z)J(w) &\sim {3\over (z-w)^3}+{J(w)\over (z-w)^2}+{\partial J(w)\over
(z-w)}\cr
J(z)Q_s(w) &\sim {Q_s(w)\over (z-w)}\cr
J(z)G(w) &\sim {-G(w)\over (z-w)}\cr
J(z)J(w) &\sim {-3\over (z-w)^2}\cr}
$$
\vfil
\eject
$$
\eqalign{Q_s(z)Q_s(w) &\sim 0\cr
Q_s(z)G(w) &\sim {-3\over (z-w)^3}+{J(w)\over (z-w)^2}+{T(w)\over (z-w)}\cr
G(z)G(w) &\sim 0}\eqno(8)
$$
Here, $Q_s(z)$ is the fermionic or supersymmetry current and has conformal
weight 1 and ghost number 1 with respect to the ghost number current (7b).
Note that with respect to (7b),
$b(z)$, $c(z)$, $\beta (z)$ and $\gamma (z)$ have ghost
numbers $-1$, $+1$, $-2$ and $+2$ respectively. The Virasoro $BRST$
charge is given by:
$$
\eqalign{Q_v &=\oint{dz :c(z)\left( T_M(z)+T_L(z)+{1\over 2}T^{gh}(z)
\right) }:\cr
&=\oint{dz :c(z)\biggl( 2\beta (z)\partial\gamma (z) +\partial\beta (z)
\gamma (z)-b(z)\partial c(z)\biggr) :} }\eqno(9)
$$
It has been argued in ref. [14] that the total $BRST$ charge in the
topological
gravity is not just $Q_v$, but $(Q_s+Q_v)$\footnote{$^1$}{This is also implicit
in the gauge transformations (Eqs.(4.5),(5.1)-(5.3)) of ref.[6].}.
It is easy to check that $Q_s$
and $Q_v$ anticommute. We note also that the cohomology class associated with
$(Q_s+Q_v)$ is isomorphic to that of $Q_s$. In fact, there exists a
homotopy operator $U$:
$$
U=:e^{-{1\over 2}\oint{dz c(z)G(z)}}:\eqno(10)
$$
(where $G(z)$ is given in (7c)) by which $(Q_s+Q_v)$ is related to $Q_s$ by
a similarity transformation, namely,
$$
U(Q_s+Q_v)U^{-1}=Q_s\eqno(11)
$$
This type of transformation was used in the context of topological strings
in ref. [15]. One can also verify that,
under the similarity
transformation (11),
the fields $b(z)$, $c(z)$, $\beta (z)$ and $\gamma (z)$ transform as:
$$
\eqalignno{Ub(z)U^{-1}&=b(z)+G(z)  &(12a)\cr
Uc(z)U^{-1}&=c(z) &(12b)\cr
U\beta (z)U^{-1}&=\beta (z) &(12c)\cr
U\gamma (z)U^{-1} &=\gamma (z)-c(z)\partial c(z)   &(12d)}
$$

By making use of our knowledge about
the cohomology class of the
$BRST$ charge for $(p,q)$ minimal model coupled to gravity, we now study the
cohomology class of $Q_s$. In particular, we look at the ground ring
generators of $c_M<1$ string theory. These are ghost number zero and
conformal weight zero operators and have the form:
$$
\eqalignno{x(z) &=:\biggl[ b(z)c(z)-{1\over 2\lambda}(i\partial\phi_M(z)
 -\partial\phi_L(z))\biggr] e^{i\lambda (\phi_M(z)-i\phi_L(z))}: &(13a)\cr
y(z) &=:\biggl[ b(z)c(z)+\lambda (i\partial\phi_M(z)+\partial\phi_L(z))
\biggr] e^{-{i\over 2\lambda}(\phi_M(z)+i\phi_L(z))}: &(13b)}
$$
Note that under the symmetries

$$
\lambda\leftrightarrow {1\over 2\lambda},~~~~~~~
\phi_M(z)\leftrightarrow -\phi_M(z)\eqno(14a)
$$
and
$$
\lambda\leftrightarrow
{1\over 2i\lambda},~~~~~~~ \phi_M(z)\leftrightarrow\phi_L(z)
\eqno(14b)
$$
(these are
symmetries of the energy-momentum tensor (1) as well as the BRST charge $Q_v$
(9)), the ground ring generators
$x(z)$ and $y(z)$ interchange their roles. For the symmetry (14a), $p$ and
$q$ get interchanged. It is easy to check that both,
$x(z)$ and $y(z)$, are in the cohomology of $(Q_s+Q_v)$. Also,
under the similarity transformation generated by (10)
 we find that both $x(z)$ and $y(z)$ remain invariant, namely,
$$
Ux(z)U^{-1} =x(z), ~~~~~~~~~~~~~~ Uy(z)U^{-1}=y(z)\eqno(15)
$$
In other words, both $x(z)$ and $y(z)$ are in the
cohomology of $Q_s$. In the case of the field $x(z)$, this fact can
be more easily seen by noticing that this ground ring generator can be
expressed in terms of the fields in the topological gravity, $i.e.$, $b(z)$,
$c(z)$, $\beta (z)$
and $\gamma (z)$, and then using the equations (12a-12d).
 In order to do that we here note the following
relations,
$$
\eqalign{:\gamma^2 (z)\beta (z): &=:\biggl[
-{1\over 2}({1\over \lambda}+3\lambda)
\partial\phi_L(z)+{i\over 2}({1\over \lambda}-3\lambda)\partial\phi_M(z)
\biggr] e^{i\lambda(\phi_M(z)-i\phi_L(z))}:\cr
\partial\gamma (z) &=:\biggl[
\lambda\partial\phi_L(z)+i\lambda\partial\phi_M(z)
\biggr] e^{i\lambda (\phi_M(z)-i\phi_L(z))}:}\eqno(16)
$$
Using (15) we can express $:\partial\phi_L(z) e^{i\lambda (\phi_M(z)-
i\phi_L(z))}:$ and $:\partial\phi_M(z) e^{i\lambda (\phi_M(z)-i\phi_L(z))}:$
in terms of $b(z)$, $c(z)$, $\beta (z)$ and $\gamma (z)$ as follows,
$$
\eqalign{:\partial\phi_L(z) e^{i\lambda (\phi_M(z)-i\phi_L(z))}: &=
-\lambda :\gamma^2 (z)\beta (z):+{1\over 2}({1\over \lambda }-3\lambda )
\partial\gamma (z)\cr
:\partial\phi_M(z) e^{i\lambda (\phi_M(z)-i\phi_L(z))}: &=
-i\lambda :\gamma^2 (z)\beta (z):-{i\over 2}({1\over \lambda }+3\lambda )
\partial\gamma (z)}\eqno(17)
$$
Substituting (17) in (13a) we find:
$$
x(z)=:b(z)c(z)\gamma (z):-:\gamma^2 (z)\beta (z):-{3\over 2}\partial\gamma (z)
\eqno(18)
$$
which is independent of $\lambda$. It should be pointed out here that the
physical operators in the topological gravity are expressed as $\gamma_0^n
\sim (\partial\gamma)^n$ [7]. The ground ring generator $x(z)$ in (18)
differs from $\partial\gamma$ by $Q_s$-exact terms, since $:b(z)c(z)\gamma
(z) - \gamma^2(z)\beta(z): = $\hfill

\noindent$\{Q_s, -:c(z)\gamma(z)\beta(z):\}$.
It is interesting to note the similarity between (18) and the local
Lorentz ghost expression (3.12) in ref. [7].

On the other hand, we also find that the other
ground ring generator $y(z)$ can not be expressed locally in terms of $b(z)$,
$c(z)$, $\beta (z)$ and $\gamma (z)$ of the topological
gravity. Then, to verify equation (15) for $y(z)$, one has to use
the string representation of $G(z)$ in the homotopy operator (10) ($i.e.$,
the field $\beta (z)$ in (7c) should be expressed as given in
(4)) in order to verify (15).

 Another interesting point to note here is that there
exists another bosonization (4) using the symmetries mentioned in (14a) and
(14b).
Since in this case
$x(z)$ and $y(z)$ interchange their roles, with that bosonization
$y(z)$ can be expressed in terms of
$b(z)$, $c(z)$, $\beta (z)$ and $\gamma (z)$ and not $x(z)$.
In fact, in general, the
statements true for $x(z)$ in one bosonization will be
valid for $y(z)$ in the other bosonization. We would also
like to point out that
in $c_M<1$ string theory, the appearance of two ground ring generators $x$
and $y$ in a symmetric way can be attributed to the symmetry observed in (14).
But in the topological gravity representation, the fermionic charge $Q_s$
takes the place of the BRST charge of $c_M<1$ string theory. Since $Q_s$
does not respect this symmetry (14), it is perhaps not surprising that in one
bosonization both $x$ and $y$ can not be expressed in terms of $(b(z), c(z),
\beta(z), \gamma(z))$.

Let us study the ground ring generators $x(z)$ and $y(z)$ in more detail.
First, we note that even though both the ground ring generators are in the
cohomology of $Q_s$, both are trivial in the sense that they are $Q_s$ exact.
To be precise, we find:
$$
\eqalignno{x(z) &=\{ Q_s,\Lambda_x(z)\} &(19a)\cr
y(z) &=\{ Q_s, \Lambda_y(z)\} &(19b)}
$$
where
$$
Q_s =\oint{dz :b(z) e^{i\lambda (\phi_M(z)-i\phi_L(z))}}:\eqno(20)
$$
and
$$
\eqalignno{\Lambda_x(z) &=({1\over 2\lambda}-\lambda)c(z)\partial\phi_L(z)-
i({1\over 2\lambda}+\lambda)c(z)\partial\phi_M(z)\cr
&\qquad\qquad +k\bigl[\partial c(z)-\lambda
c(z)\partial\phi_L(z)-i\lambda c(z)\partial\phi_M(z)\bigr]\cr
&\equiv ({1\over 2\lambda}-\lambda)c(z)\partial\phi_L(z)-
i({1\over 2\lambda}+\lambda)c(z)\partial\phi_M(z)+k\psi(z)
&(21)\cr
{}\cr
\Lambda_y (z) &=:c(z)e^{-i(\lambda +{1\over 2\lambda})\phi_M(z)-(\lambda -
{1\over 2\lambda})\phi_L(z)}: &(22)}
$$
Here, $k$ is an arbitrary constant and we have defined $\psi (z)\equiv
\partial c(z)-\lambda c(z)\partial\phi_L(z)-i\lambda c(z)\partial\phi_M(z)$.

In topological gauge theories it is usual to consider the equivariant
cohomology [16-18], otherwise the theories
become empty. This means that we only
allow states in the Hilbert space which are annihilated by the antighost
zero-mode $b_0$. Because of the similarity transformation (12a), the new
equivariance condition would be $(b_0 +G_0)|state>=0$. $(b_0 +G_0)$ will
act on an operator $\cal O$ as,
$$
\oint{dz~z(b(z)+G(z)){\cal O}(0)}\eqno(23)
$$
Using (23) we find:
$$
(b_0+G_0)\Lambda_x (0)=k\eqno(24)
$$
and
$$
(b_0+G_0)\Lambda_y (0)=0\eqno(25)
$$
This shows that the ground ring generator $y(z)$ is trivial even in the
equivariant cohomology of $Q_s$ and $x(z)$ might seem non-trivial when
$k\neq 0$. But, it is easy to check that $Q_s$ acts trivially on
$\psi (z)$, since:
$$
\psi (z)=[Q_v,\ -\lambda (\phi_L(z)+i\phi_M(z))]\eqno(26a)
$$
and
$$
[Q_s,\ \lambda (\phi_L(z)+i\phi_M(z))]=0\eqno(26b)
$$
Note that the operator $\psi (z)$ is not $Q_s$-exact and therefore, $\psi (z)
$ is a non-trivial physical state in the topological gravity. It is clear
that this operator can not be expressed locally in terms of $b(z)$, $c(z)$,
$\beta (z)$ and $\gamma (z)$. $\psi (z)$ is the analog of the operator
`$a(z)$' discussed
in ref. [19]. We just like to emphasize that even when $k \neq 0$, the ground
ring generator $x(z)$ is trivial in the equivariant cohomology as long as
$\Lambda_x (z)$ can be split into the form as given in (21). This will no
longer be possible if we try to express $\Lambda_x (z)$
in terms of $b(z)$, $c(z)$,
$\beta (z)$ and $\gamma (z)$. We find that, since
$$
:\gamma (z)\beta (z):=-{1\over 2}(\lambda +{1\over \lambda})\partial\phi_L(z)-
{i\over 2}(\lambda -{1\over\lambda})\partial\phi_M(z)\eqno(27)
$$
only for $k=-{3\over 2}$ one has
$$
\Lambda_x (z)=-{3\over 2}\partial c(z)-:c(z)\gamma (z)\beta(z):\eqno(28)
$$
Again, we note the similarity between (28) and the anticommuting local
Lorentz ghost expression in the equation (3.3) in ref. [7]. So, in the
topological gravity representation, one can not split $\Lambda_x(z)$ as
in (21) and, since $k\neq 0$, we will find the ground ring generator
$x(z)$ in the equivariant cohomology of $Q_s$.

The other ground ring generator $y(z)$ on the usual $SL(2,\bf C)$ invariant
vacuum is, as noted before, trivial on the equivariant cohomology. We,
however, find it non-trivial on a ``picture-changed'' [20] vacuum. A vacuum
with picture charge $q_p$ is defined as:
$$
|q_p>=\lim_{z\rightarrow 0}:e^{q_p[ {i\over 2}(\lambda -{1\over\lambda})
\phi_M(z)+{1\over2}(\lambda +{1\over\lambda})\phi_L(z)]}:|0>\eqno(29)
$$
where $|0>$ is the $SL(2,\bf C)$ invariant vacuum. It is easy to check
that $|q_p>$ is
invariant under $Q_s$ for $q_p\leq 0$. In terms of modes, $|q_p>$
satisfy:
$$
\eqalignno{b_n|q_p> &=0 ~~~~~~~~ n\geq {-1}\cr
c_n|q_p> &=0 ~~~~~~~~ n\geq 2\cr
\beta_n |q_p> &=0 ~~~~~~~~ n\geq {-1-q_p}\cr
\gamma_n |q_p> &=0 ~~~~~~~~ n\geq {2+q_p} &(30)}
$$
We would like to mention that the ground ring generator $x(z)$ is well-defined
on $|q_p>$
as long as $q_p$ is an integer, but for $q_p<0$, $x(z)$ is trivial on $|q_p>$.
For the other ground ring generator $y(z)$, it is well-defined only when $q_p$
is even and if we want to have a $Q_s$ invariant vacuum, $q_p$ has to be
negative. In fact, for
all $q_p<-2$, $y(z)$ is trivial on $|q_p>$. So, $y(z)$ is
non-trivial only in the `$-2$' picture. It is not difficult to compute $y(z)$
in this picture, and has the form:
$$
y(z)|_{q_p=-2} =:2 \ e^{iQ_M\phi_M(z)+iQ_L\phi_L(z)}:\eqno(31)
$$
where $Q_M$ and $Q_L$ are as given in (3).
But, again, we find that $y(z)|_{q_p=-2}$ is $Q_s$-exact. In other words,
$$
y(z)|_{q_p=-2} =\{ Q_s,\Lambda_y (z) |_{q_p=-2}\}\eqno(32)
$$
where,
$$
\eqalign{\Lambda_y (z)|_{q_p=-2} =&:\biggl[(2-\lambda (\alpha_1+i\alpha_2))
\partial c(z)+
\alpha_1 c(z)\partial\phi_L(z)+\alpha_2 c(z)\partial\phi_M(z)\biggr]\cr
& \qquad\qquad \times e^{i({1\over 2\lambda}-2\lambda)\phi_M(z)-
({1\over 2\lambda}+2\lambda)\phi_L (z)}:}\eqno(33)
$$
with $\alpha_1$ and $\alpha_2$ some arbitrary constants.
Now, $\Lambda_y (z)|_{q_p=-2}$ is no-longer annihilated by $(b_0+G_0)$. This
clearly indicates that $y(z)$ is also in the equivariant cohomology but in
$q_p=-2$ picture. However, as we have pointed out above, $y(z)$ can not be
expressed in terms of the fields $\beta (z)$, $\gamma (z)$, $b(z)$ and $c(z)$
and, in this sense,
the significance of $y(z)$ in the context of topological gravity remains
unclear.

Finally, we would like to mention here, that one can use the topological
descent equations to obtain the one-form physical operators from the zero-form
operators as [19],
$$
\Omega^{(1)} (0) = \left[\oint dz G(z), \Omega^{(0)} (0)\right]\eqno(34)
$$
where $G(z)$ is as given in (7c). In particular, we find for $x(z)$, $y(z)$
and $\psi(z)$ the corresponding one-forms are
$$
\eqalignno{\Omega_x^{(1)}(z) &= :c(z)\left(T_M(z) + T_L(z)+ {1\over 2}T^{gh}(z)
\right): + ({1\over \lambda} + \lambda)\partial(c(z)\partial \phi_L(z))\cr
&\qquad\qquad\qquad -i ({1\over \lambda}-\lambda)\partial(c(z)
\partial\phi_M(z)) -{3\over 2}\partial^2 c(z) &(35a)\cr
\Omega_y^{(1)}(z) &= :\partial \left(c(z)\ e^{-i(\lambda + {1\over 2\lambda})
\phi_M
(z) -(\lambda -{1\over 2\lambda})\phi_L(z)}\right): &(35b)\cr
\Omega_\psi^{(1)}(z) &= :\partial\left(c(z)\partial c(z)\ e^{-i \lambda
(\phi_M(z) -i\phi_L(z))}\right): &(35c)\cr}
$$
It is quite intriguing to note that $\Omega_x^{(1)}(z)$ is precisely the
modified BRST current used in equation (8) of ref. [11] for $a_3 =-{3\over 2}
$.

 From the above analysis, we conclude, that although any $(p, q)$ minimal
model coupled to $2d$ gravity has topological gravity representation, it
is not clear how to recover the full spectrum of $c_M<1$ string theory in
this representation. The reason for this may be traced back to the symmetry
charge $(Q_s+Q_v)$ (not just $Q_v$, the string BRST charge) associated with
the topological gravity of ref.[6]. Since the cohomology class of $(Q_s+Q_v)$
is different from the cohomology class of $Q_v$ itself, it is in a
sense natural that we do not get the whole spectrum of $c_M<1$ string theory
in the topological gravity representation. Interestingly, we have noticed
that in the gauge fixed form of the topological gravity formulation of
ref.[6], there exists a twisted $N=2$ superconformal algebra where the
topological charge is precisely the BRST charge of $c_M<1$ string theory.
How this information would help us to recover the full spectrum of $c_M<1$
string theory is not clear at this moment. Recently, however, we have seen
that $c_M<1$ string theory can be regarded as a constrained topological
sigma model [21] in analogy with the similar result for $c_M=1$ string
theory found in ref.[22] and recovered the spectrum of $c_M<1$ string theory.

To conclude, we have shown the presence of a new topological algebra
other than the one found in ref. [10]
 in ordinary $2d$ gravity coupled to $(p,q)$ minimal model. This
algebra was known in the topological gravity formulation of Labastida,
Pernici and Witten. This, therefore, reveals a topological gravity
structure in general $c_M <1$ string theory. We have also studied the
cohomology class associated with the topological charge $Q_s$ using the
knowledge of the cohomology class associated with the Virasoro $BRST$
charge. In particular, we have shown that one of the ground ring generators
$x(z)$ does not belong to the equivariant cohomology of $Q_s$ in the string
representation, but belong to this cohomology in the topological gravity
representation. This kind of ambiguity is not present in the other ground
ring generator $y(z)$. In fact, for $y(z)$, we found that it is trivial
in the equivariant cohomology if we consider it in the representation
$q_p =0$. But for $q_p=-2$, it is non-trivial. We also noted that all the
observations made for $x(z)$ and $y(z)$ are only true for the particular
bosonization (4). In the other bosonization (14), $x(z)$ and $y(z)$ will
be interchanged. We, however, like to mention here that we do not get the
usual restrictions [13] on the ground ring generators of the form $x^{p-1}
= y^{q-1} = 0$, which were imposed in $c_M<1$ string theory from the
requirements that the matter momenta of the physical states should lie inside
the Kac-table. To understand how to obtain the tachyons as well as the higher
ghost number operators  [23,24]
of $c_M < 1$ string cohomology  in the topological gravity formulation we
have obtained, requires further study in this direction.

\vskip 1cm
\noindent{\titlea Acknowledgements:}
\medskip
We would like to thank E. Bergshoeff,  H. J. Boonstra,
J. M. F. Labastida, A. V. Ramallo and M. de Roo for
discussions at various stages of the
work. We also thank K. Thielemans for providing us his latest version of the
Mathematica package OPEconf and ref.[25] which was used in some OPE
computations. The
work of S. R. was performed as part of the research program of the ``Stichting
voor Fundamenteel Onderzoek der Materie'' (FOM). The work of P. M. Ll. is
supported by the ``Human Capital and Mobility Program'' of the European
Community.
\vfill
\eject

\noindent{\titlea References:}
\item{1.}E. Witten, Nucl. Phys. B314 (1990) 281; Surv. Diff. Geom. 1 (1991)
243.
\item{2.} K. Li, Nucl. Phys. B354 (1991) 711,725.
\item{3.} E. Witten, Comm. Math. Phys. 118 (1988) 411; Nucl. Phys. B340 (1990)
281.
\item{4.} T. Eguchi and S. -K. Yang, Mod. Phys. Lett. A4 (1990) 1693.
\item{5.} J. Distler, Nucl. Phys. B342 (1990) 523.
\item{6.} J.M.F. Labastida, M. Pernici and E. Witten,
Nucl. Phys. B310 (1988) 611.
\item{7.} E. Verlinde and H. Verlinde, Nucl. Phys. B348 (1991) 457.
\item{8.} R. Dijkgraaf, E. Verlinde and H. Verlinde, Preprint PUPT-1217,
IASSNS-HEP-90/80 (1990).
\item{9.} B. Gato-Rivera and A. Semikhatov, Phys. Lett. B288 (1992) 295.
\item{10.} M. Bershadsky, W. Lerche, D. Nemeschansky and N.P. Warner,
Nucl. Phys. B401 (1993) 304.
\item{11.} S. Panda and S. Roy, Phys. Lett. B317 (1993) 533.
\item{12.} E. Witten, Nucl. Phys. B373 (1992) 187.
\item{13.} D. Kutasov, E. Martinec and N. Seiberg, Phys. Lett. B276 (1992) 437.
\item{14.} J. Distler and P. Nelson, Phys. Rev. Lett. 66 (1991) 1955.
\item{15.} T. Eguchi, H. Kanno, Y. Yamada and S. -K. Yang, Phys. Lett. B305
(1993) 235.
\item{16.} E. Witten, Comm. Math. Phys. 117 (1988) 353.
\item{17.} S. Ouvry, R. Stora and P. van Baal, Phys. Lett. B220 (1989) 159.
\item{18.} J. Distler and P. Nelson, Comm. Math. Phys. 138 (1991) 273;
Nucl. Phys. B366 (1991) 255.
\item{19.} E. Witten and B. Zwiebach, Nucl. Phys. B377 (1992) 55.
\item{20.} D. Friedan, E. Martinec and S. Shenker, Nucl. Phys. B271 (1986) 93.
\item{21.} P. M. Llatas and S. Roy, preprint to appear.
\item{22.} H. Ishikawa and M. Kato, preprint UT-Komaba/93-7, hep-th/9304039
(revised Dec. '93).
\item{23.} H. Kanno and M.H. Sarmadi, Int. Jour. Mod. Phys. A9 (1994), 39.
\item{24.} S. Panda and S. Roy, Phys. Lett. B306 (1993) 252.
\item{25.} K. Thielemans, Int. Jour. Mod. Phys. C2 (1991) 787.

\end